\documentclass[aps,pra,twocolumn,groupedaddress,showkeys]{revtex4}
\usepackage{graphicx}% Include figure files
\usepackage{dcolumn}% Align table columns on decimal point
\usepackage{bm}% bold math
\usepackage{amsmath}
\usepackage{amssymb}
\usepackage{color} %\textcolor[rgb]{0,0,1}{text}
\usepackage{slashed}

\begin{document}

\title{Arrival time from Hamiltonian with non-hermitian boundary term}

\author{Tajron Juri\'c}\email{tjuric@irb.hr} 
\author{Hrvoje Nikoli\'c}\email{hnikolic@irb.hr}
\affiliation{Theoretical Physics Division, Rudjer Bo\v{s}kovi\'{c} Institute, P.O.B. 180, HR-10002 Zagreb, Croatia.}

\begin{abstract}
We develop a new method for finding the quantum probability density of arrival at the detector.
The evolution of the quantum state restricted to the region outside of the detector
is described by a restricted Hamiltonian that contains a non-hermitian boundary term.
The non-hermitian term is shown to be proportional to the flux of the probability current operator
through the boundary, which implies that the arrival probability density is equal to the 
flux of the probability current. 
\end{abstract}


\maketitle

\section{Introduction} 
 
Consider a quantum particle described by a spatially extended wave packet impinging on the detector region $D$. 
Since different parts of the packet approach $D$ at different times, 
there is an inherent quantum uncertainty about the time at which the particle arrival to $D$ will 
be detected. The arrival time problem is to make a theoretical prediction for the probability distribution 
${\cal P}_{\rm arr}(t)$ that the arrival will be detected at time $t$. Remarkably, there are many different 
theoretical approaches to that problem that make different measurable predictions, 
for reviews see \cite{muga-physrep,V1.10,V2.4,maccone},
and it is not clear, neither theoretically nor experimentally, which approach is correct.   

%\textcolor[rgb]{0,0,1}{
In general, quantum mechanics makes unambiguous probabilistic predictions for various phenomena, 
so why the arrival time problem is a problem at all? The point is that quantum mechanics 
makes unambiguous probabilistic predictions for measurements of observables represented by
self-adjoint operators, while time, in the usual formulation of 
quantum mechanics, is not an observable in that sense. Time is a classical parameter, not a 
quantum operator, so, by starting from general axioms of quantum theory, it is not immediately 
clear how to make quantum probabilistic predictions associated with measurement of time.
In particular, one does not know at which time a quantum event, such as a particle detection, 
will happen, so one wants to use quantum mechanics to compute a probability that the event will happen 
at a given time. The problem is, how to compute this probability, when the time is not an operator?
The arrival time problem is the simplest version of this problem, where the quantum event 
is taken to be the particle arrival to the detector, or more operationally, a click in the detector 
which happens when the particle arrives.
%}

%\textcolor[rgb]{0,0,1}{
One class of possibilities (see e.g. \cite{muga-physrep,maccone} and references therein) 
is to reformulate quantum mechanics such that time is treated as an operator, 
the problem with such approaches is that they may require a radical reformulation 
of general principles of quantum mechanics, which makes them rather controversial. 
There are also axiomatic approaches by Kijowski and others (see e.g. \cite{muga-physrep,maccone} and references therein)
that postulate axioms for the arrival time distribution, the problem is that those axioms seem somewhat {\it ad hoc}
because they cannot be derived from standard axioms of quanum mechanics. 
Another class of possibilities (see e.g. \cite{maccone} and references therein)
are semi-classical approaches, the problem is that they also seem too {\it ad hoc}
and lack a deeper understanding of the problem. 
%}

%\textcolor[rgb]{0,0,1}{
Yet another class  
\cite{kijowski,delgado-muga,tumulka1,nik_tajron2,leavens,leavens_book,durrtime1,durrtime2}
of approaches to the arrival time problem
%}
predicts that ${\cal P}_{\rm arr}(t)$ is given by the flux of the probability current. 
Within this class, some approaches are based on standard quantum mechanics (QM)
\cite{kijowski,delgado-muga,tumulka1,nik_tajron2}, while others are based on Bohmian formulation of QM
in terms of particle trajectories \cite{leavens,leavens_book,durrtime1,durrtime2}.
In this paper we present one new approach to the arrival time problem, based on standard QM, which 
confirms that the arrival time distribution is given by the flux of the probability current. Given that 
it is not generally accepted in the community that the arrival time distribution should be given
by the flux of the probability current, we believe that it is valuable to present one more independent 
theoretical evidence that it is indeed so.
 
The approach in this paper is partially inspired by the approach in \cite{nik_tajron2}, but is motivated 
with a goal to avoid some mathematical subtleties 
that appeared in that work. The approach in \cite{nik_tajron2},
which arose from the development of earlier ideas in \cite{jur-nik,nik_tajron}, is based on time evolution governed by
a projected Hamiltonian $\overline{H}=\bar{\pi}H\bar{\pi}$, where $\bar{\pi}$ is the projector to the region
${\bar D}$ defined as the complement of the detector region $D$. 
The mathematical subtleties appear because, 
in the position representation, 
$\bar{\pi}$ is represented by a characteristic function with a discontinuity at the boundary of ${\bar D}$, 
which leads to ambiguities when the second-derivative operator appearing in $H$ acts on a function with a 
discontinuity.
The goal of this paper is to develop a formalism based on an alternative definition of $\overline{H}$
that, at the same time, captures the same physics as $\overline{H}$ in \cite{jur-nik,nik_tajron,nik_tajron2},
but uses a different mathematical definition of $\overline{H}$ so that the mathematical difficulties appearing in 
\cite{nik_tajron,nik_tajron2} are avoided.

The paper is organized as follows.
In Sec.~\ref{SECwf} we start by defining the notion of restricted wave function, which 
in ${\bar D}$ coincides with the full wave function, but vanishes outside of ${\bar D}$ 
where the full wave function, in general, does not vanish.
This implies that the norm of the restricted wave function,
in general, is not conserved in time. 
Then in Sec.~\ref{SECH} we define the restricted Hamiltonian $\overline{H}$ 
that governs the evolution of the restricted wave function. It turns out that
the restricted Hamiltonian is non-trivial at the boundary of ${\bar D}$, because it
has a non-hermitian boundary term proportional to the flux of the probability current operator,
which accounts for non-conservation of the norm of the restricted wave function.
In Sec.~\ref{SECarrival} we explain how this non-conservation of the norm implies that the arrival time distribution 
is equal to the flux of the probability current.
The conclusions are drawn in Sec.~\ref{SECconclusion}, 
and in the Appendix an alternative derivation of the non-hermitian part is presented.

\section{Restricted wave function}
\label{SECwf}

Let us start with an elementary warmup needed to establish the notation. 
Consider a particle moving in a  3-dimensional space $\mathbb{R}^3$. It is described by a wave function 
\begin{equation}
 \psi({\bf x},t) =\langle {\bf x} |\psi(t)\rangle ,
\end{equation}
where $|\psi(t)\rangle$ is the state in the Hilbert space $\mathcal{H}$ denoted in the Dirac's ``bra-ket'' formalism. 
The $|\psi(t)\rangle$ and $\psi({\bf x},t)$ 
satisfy the respective Schr\"odinger equations
\begin{equation}\label{wf2}
 H |\psi(t)\rangle = i\partial_t |\psi(t)\rangle , \;\;\; \hat{H}\psi({\bf x},t)=i\partial_t\psi({\bf x},t) ,
\end{equation}
where $H$ is an abstract operator, while $\hat{H}$ is its coordinate representation given by
a concrete derivative operator
\begin{equation}\label{e1}
H=\frac{{\bf p}^2}{2m}+V({\bf x}) ,  \;\;\; \hat{H}= -\frac{\mbox{\boldmath $\nabla$}^2}{2m}+V({\bf x}) ,  
\end{equation}
and we work in units $\hbar=1$. The relation between $H$ and $\hat{H}$ can be expressed as 
\begin{equation}
 \langle{\bf x}|H|\psi(t)\rangle = \hat{H} \psi({\bf x},t) .
\end{equation}
 
Now, after this warmup, let us divide the full space $\mathbb{R}^3$ 
into a detector region $D$ and its complement $\bar{D}$, so that $D\cup\bar{D}=\mathbb{R}^3$,
$D\cap\bar{D}=\emptyset$. 
Physically, such a division is motivated with the goal to study the arrival of the particle 
to the detector.
We define the {\em restricted wave function} $\psi_{\bar D}({\bf x},t)$ as
\begin{equation}
 \psi_{\bar D}({\bf x},t) \equiv\left\{ 
\begin{array}{cl}
 \displaystyle \psi({\bf x},t) & \;\; {\rm for} \;\; {\bf x}\in \bar{D}  \\
 0 & \;\; {\rm for} \;\; {\bf x}\notin \bar{D} .
\end{array}
\right.
\end{equation}
It can be expressed as 
\begin{equation}
 \psi_{\bar D}({\bf x},t)=\langle {\bf x}|\bar{\pi}|\psi(t)\rangle ,
\end{equation}
where 
\begin{equation}\label{barpi}
  \bar{\pi}=\int_{\bar D} d^3x\,|{\bf x}\rangle\langle{\bf x}| 
\end{equation}
is the projector to ${\bar D}$. Thus we see that the restricted wave function can also be expressed as
\begin{equation}
 \psi_{\bar D}({\bf x},t)=\langle {\bf x}|\psi_{\bar D}(t)\rangle ,
\end{equation}
where
\begin{equation}
 |\psi_{\bar D}(t)\rangle = \bar{\pi}|\psi(t)\rangle .
\end{equation}

Since $\psi_{\bar D}({\bf x},t)$ coincides with $\psi({\bf x},t)$ in $\bar{D}$, but vanishes outside of $\bar{D}$,
we have
\begin{equation}\label{ineq}
\int _{\mathbb{R}^3} d^3x \, |\psi_{\bar D}({\bf x},t)|^2  \leq \int _{\mathbb{R}^3} d^3x \, |\psi({\bf x},t)|^2 .
\end{equation}
In fact, since we assume that $\psi({\bf x},t)$ is a travelling wave packet (rather than a stationary state),
one expects that the left-hand side depends on time $t$, despite the fact that the right-hand side is time independent.
This means that the norm of $\psi_{\bar D}({\bf x},t)$ is expected to change with time, i.e., that the norm 
$\left\|\psi_{\bar D}(t)\right\|=\langle\psi_{\bar D}(t) |\psi_{\bar D}(t)\rangle^{1/2}$ is time-dependent. 
Consequently, one expects that the evolution 
of $|\psi_{\bar D}(t)\rangle$ is not unitary. By contrast, the evolution of the full state 
$|\psi(t)\rangle$ is unitary.

%\textcolor[rgb]{0,0,1}{
The restricted wave function can also be interpreted in terms of wave function collapse. 
When a part of the wave function enters the detector region $D$, there is a non-zero probability 
that the detector will detect the particle, i.e., that the wave function will collapse to the region 
$D$. But there is also a probability that the detector will not detect the particle, 
in which case we know that the particle is still outside of the detector region 
$D$, so the wave function collapses to the region $\bar{D}$. Thus the restricted wave function  
can be interpreted as the collapsed wave function, corresponding to a negative 
outcome of measurement by the detector. For more details of such an interpretation see 
\cite{nik_tajron2}. 
%}

%\textcolor[rgb]{0,0,1}{
From a theoretical point of view, the exact specification of the detector region $D$ is 
in our analysis still somewhat ambiguous. In principle, one could specify it by a more detailed model.
But in practice, we believe that an experimental approach would be more fruitful.
One could take an actual detector and impinge on it particles with wave functions which are
very narrow in the position space, so that particles are effectively ``classical" in the sense
that their arrival time can be predicted as a classical deterministic event.
In this way, one can determine the relevant detector region $D$ experimentally, 
for the specific detector at hand. After that, once $D$ is known, one can make non-trivial theoretical analysis with 
``truly quantum" non-narrow wave functions.
%}   

\section{Restricted Hamiltonian}
\label{SECH}

The full state evolves with time as $|\psi(t)\rangle = e^{-iHt}|\psi(0)\rangle$. This evolution is unitary 
because the Hamiltonian $H$ is hermitian. On the other hand, since one expects that the evolution 
of the restricted state $|\psi_{\bar D}(t)\rangle$ is not unitary, its evolution can be described as 
\begin{equation}\label{r0}
|\psi_{\bar D}(t)\rangle=e^{-i\overline{H}t}|\psi_{\bar D}(0)\rangle ,
\end{equation} 
where $\overline{H}$ is expected to be some 
{\em non-hermitian} operator. Our goal is to find an explicit expression for $\overline{H}$.

Heuristically, since $\psi_{\bar D}({\bf x},t)$ and $\psi({\bf x},t)$ coincide for ${\bf x}\in\bar{D}$, 
the coordinate representation $\hat{\overline{H}}$ must coincide with the derivative operator $\hat{H}$ in (\ref{e1})
for ${\bf x}\in\bar{D}$. Likewise, since $\psi_{\bar D}({\bf x},t)=0$ for ${\bf x}\notin\bar{D}$,
the operator $\hat{\overline{H}}$ can be taken as the trivial zero operator for ${\bf x}\notin\bar{D}$.
However, a particular care should be taken about definition of $\hat{\overline{H}}$ at the {\em boundary}
of $\bar{D}$, which is the only place where subtleties in the definition of $\hat{\overline{H}}$
may be expected. For that purpose we find more convenient to work with the abstract $\overline{H}$ operator,
rather than its coordinate representation $\hat{\overline{H}}$. Thus, since the arbitrary matrix element of $H$ is 
\begin{equation}\label{e2}
\langle\psi_b|H|\psi_a\rangle = \int_{\mathbb{R}^3} d^3x \, \psi^*_b({\bf x}) 
\left[  -\frac{\mbox{\boldmath $\nabla$}^2}{2m}+V({\bf x}) \right] \psi_a({\bf x}) ,
\end{equation}
we postulate that the arbitrary matrix element of $\overline{H}$ is
\begin{equation}\label{e3}
\langle\psi_b|\overline{H}|\psi_a\rangle \equiv \int_{\bar D} d^3x \, \psi^*_b({\bf x}) 
\left[  -\frac{\mbox{\boldmath $\nabla$}^2}{2m}+V({\bf x}) \right] \psi_a({\bf x}) ,
\end{equation}
which has the same form as (\ref{e2}), except that the integration region is restricted from $\mathbb{R}^3$ to ${\bar D}$.
Hence we refer to $\overline{H}$ as the {\em restricted Hamiltonian}. 
The goal now is to find the explicit operator representation of $\overline{H}$, analogous to $H$ in (\ref{e1}).

We first write (\ref{e3}) as
\begin{equation}
\langle\psi_b|\overline{H}|\psi_a\rangle = 
 -\frac{1}{2m} \int_{\bar D} d^3x\, \psi_b^* \mbox{\boldmath $\nabla$}^2 \psi_a + \overline{V}_{ba} ,
\end{equation}  
where 
\begin{equation}
\overline{V}_{ba} = \int_{\bar D} d^3x\,\psi_b^*V\psi_a 
= \int_{\bar D} d^3x\, V\psi_b^*\psi_a
= \int_{\bar D} d^3x\, \psi_b^*\psi_a V .
\end{equation}
Hence partial integration and the Gauss theorem give
\begin{eqnarray}\label{e5}
\langle\psi_b|\overline{H}|\psi_a\rangle &=& 
 \frac{1}{2m} \int_{\bar D} d^3x\, (\mbox{\boldmath $\nabla$}\psi_b)^* (\mbox{\boldmath $\nabla$} \psi_a) 
\nonumber \\
& & -\frac{1}{2m} \int_{\bar D} d^3x\, \mbox{\boldmath $\nabla$} (\psi_b^*\mbox{\boldmath $\nabla$} \psi_a)
 + \overline{V}_{ba} 
\nonumber \\
&=& 
 \frac{1}{2m} \int_{\bar D} d^3x\, (\mbox{\boldmath $\nabla$}\psi_b)^* (\mbox{\boldmath $\nabla$} \psi_a)
\nonumber \\
& & -\frac{1}{2m} \int_{\partial\bar D} d{\bf S}\cdot (\psi_b^*\mbox{\boldmath $\nabla$} \psi_a) 
+ \overline{V}_{ba} ,
\end{eqnarray}
where $\partial\bar D$ is the boundary of $\bar D$ and $d{\bf S}$ is the area element directed outwards from $\bar D$.
Next we use the identities
\begin{eqnarray}\label{deff1}
 & \psi_a({\bf x})=\langle{\bf x}|\psi_a\rangle, \;\;\; \psi_b^*({\bf x})=\langle\psi_b|{\bf x}\rangle, &
\nonumber \\
& -i\mbox{\boldmath $\nabla$} \psi_a({\bf x}) = \langle{\bf x}|{\bf p}|\psi_a\rangle , \;\;\;
i\mbox{\boldmath $\nabla$} \psi_b^*({\bf x}) = \langle \psi_b|{\bf p}|{\bf x} \rangle , &
\end{eqnarray}
implying that (\ref{e5}) can be written as 
\begin{eqnarray}\label{e7}
 \langle\psi_b|\overline{H}|\psi_a\rangle &=& 
 \frac{1}{2m}\langle\psi_b |{\bf p}\bar{\pi}{\bf p}|\psi_a\rangle
\nonumber \\
& & -\frac{i}{2m} \int_{\partial\bar D} d{\bf S}\cdot \langle\psi_b |{\bf x}\rangle\langle{\bf x}|{\bf p}|\psi_a\rangle
\nonumber \\
& & +\langle\psi_b |\bar{\pi}V|\psi_a\rangle .
\end{eqnarray}
% where
% \begin{equation}
%   \bar{\pi}=\int_{\bar D} d^3x\,|{\bf x}\rangle\langle{\bf x}| 
% \end{equation}
% is the projector to ${\bar D}$.
Since $\bar{\pi}$ commutes with $V$, the last term in (\ref{e7}) can also be written in alternative forms
\begin{equation}
\langle\psi_b |\bar{\pi}V|\psi_a\rangle=\langle\psi_b |V\bar{\pi}|\psi_a\rangle
=\langle\psi_b |\bar{\pi}V\bar{\pi}|\psi_a\rangle ,
\end{equation}
thus we see that $\overline{H}$ can be written in the operator form
\begin{equation}\label{Hbar1}
 \overline{H} = \frac{{\bf p}\bar{\pi}{\bf p}}{2m} 
-\frac{i}{2m} \int_{\partial\bar D} d{\bf S}\cdot |{\bf x}\rangle\langle{\bf x}|{\bf p}
+\bar{\pi}V\bar{\pi} .
\end{equation}
The hermitian conjugation gives
\begin{equation}\label{Hbar1_dagger}
\overline{H}^{\dagger} = \frac{{\bf p}\bar{\pi}{\bf p}}{2m} 
+\frac{i}{2m} \int_{\partial\bar D} d{\bf S}\cdot {\bf p}|{\bf x}\rangle\langle{\bf x}|
+\bar{\pi}V\bar{\pi} ,
\end{equation}
thus we see that the first and the last term are hermitian operators, but that the middle term is not.
This shows that the restricted Hamiltonian $\overline{H}$ is not a hermitian operator, 
owing to the boundary term.

To better isolate the source of non-hermiticity it is useful to write $\overline{H}$ as
\begin{equation}\label{e4}
 \overline{H} = \frac{\overline{H}+\overline{H}^{\dagger}}{2} + \frac{\overline{H}-\overline{H}^{\dagger}}{2} ,
\end{equation}
which is convenient because the first term is manifestly hermitian and the second term manifestly anti-hermitian.
From (\ref{Hbar1}) and (\ref{Hbar1_dagger}) we
see that the two terms in (\ref{e4}) can be written as
\begin{eqnarray}\label{H-H}
 & \displaystyle\frac{\overline{H}+\overline{H}^{\dagger}}{2} = \frac{{\bf p}\bar{\pi}{\bf p}}{2m} 
+ \frac{1}{2} \int_{\partial\bar D} d{\bf S}\cdot {\bf K} + \bar{\pi}V\bar{\pi} ,
\nonumber \\
& \displaystyle\frac{\overline{H}-\overline{H}^{\dagger}}{2} = -\frac{i}{2} \int_{\partial\bar D} d{\bf S}\cdot {\bf J} , &
\end{eqnarray}
where 
\begin{equation}\label{deff2}
 {\bf K}({\bf x})=\frac{i[{\bf p},|{\bf x}\rangle\langle{\bf x}|]}{2m} , \;\;\; 
{\bf J}({\bf x})=\frac{\{{\bf p},|{\bf x}\rangle\langle{\bf x}|\}}{2m} 
\end{equation}
are hermitian operators, $[A,B]=AB-BA$ denotes a commutator, and $\{A,B\}=AB+BA$ denotes an anti-commutator. 
Thus (\ref{e4}) can be written in the final form
\begin{equation}\label{Hbar}
 \overline{H} = \left[\frac{{\bf p}\bar{\pi}{\bf p}}{2m} + \bar{\pi}V\bar{\pi} \right]
+ \frac{1}{2} \int_{\partial\bar D} d{\bf S}\cdot {\bf K} -\frac{i}{2} \int_{\partial\bar D} d{\bf S}\cdot {\bf J} .
\end{equation}
The first term (namely, the term in square brackets) 
is hermitian and does not depend on the boundary.
It is non-negative, provided that $V$ is non-negative. 
The second term (namely, the term involving ${\bf K}$) is a hermitian, but not non-negative, boundary term.
The last term (namely, the term involving ${\bf J}$) is an anti-hermitian boundary term.

%\textcolor[rgb]{0,0,1}{
Eq.~(\ref{Hbar}) is the main new result of this paper, so let us discuss its significance qualitatively.
While the full Hamiltonian (\ref{e1}) describes evolution of the full wave function everywhere in the full
3-dimensional space, (\ref{Hbar}) is the restricted Hamiltonian describing evolution of the restricted wave function, namely the part of wave function defined only on the 3-dimensional region $\bar{D}$.
The $\bar{\pi}$ is the projector to the region $\bar{D}$, so the term in square brackets in (\ref{Hbar})
is just the projected version of (\ref{e1}). The projector $\bar{\pi}$ commutes with 
$V=V({\bf x})$, but does not commute with $\bf{p}$. Hence the potential energy term 
$\bar{\pi}V\bar{\pi}$ can also be written as $\bar{\pi}V$ or $V\bar{\pi}$, but the kinetic energy term
proportional to ${\bf p}\bar{\pi}{\bf p}$ must we written in that form, and not e.g. 
as $\bar{\pi}{\bf p}\bar{\pi}{\bf p}\bar{\pi}$ or ${\bf p}\bar{\pi}{\bf p}\bar{\pi}$.
The commutator $[\bar{\pi},{\bf p}]$, in the ${\bf x}$-representation, is proportional to a 
Dirac $\delta$-function on the boundary of $\bar{D}$, so replacing ${\bf p}\bar{\pi}{\bf p}$ with
$\bar{\pi}{\bf p}\bar{\pi}{\bf p}\bar{\pi}$ or ${\bf p}\bar{\pi}{\bf p}\bar{\pi}$
would produce spurious boundary terms. In (\ref{Hbar}) all boundary terms are represented explicitly 
and unambiguously, without $\delta$-functions, as surface integrals over the boundary  
$\partial\bar{D}$ of $\bar{D}$. The most important feature of the boundary term is the fact that 
it contains an anti-hermitian part involving ${\bf J}$, the physical significance of which we discuss in more detail in the rest of the paper. 
%}

Note that the mean value of ${\bf J}$ is (see also \cite{cohen-t})
\begin{eqnarray}\label{curr}
 {\bf j}({\bf x}) & \equiv & \langle\psi|{\bf J}({\bf x})|\psi\rangle
\nonumber \\ 
 & = & \frac{-i}{2m} [ \psi^*({\bf x}) \mbox{\boldmath $\nabla$}\psi({\bf x}) 
- (\mbox{\boldmath $\nabla$}\psi^*({\bf x})) \psi({\bf x})] , 
\end{eqnarray}
which is the standard probability current in quantum mechanics. 
For that reason, we refer to ${\bf J}$ as the {\em probability current operator}.

As we said, the fact that $\overline{H}$ in (\ref{r0}) is not hermitian implies
that the norm 
\begin{equation}\label{norm}
\langle\psi_{\bar{D}}(t)|\psi_{\bar{D}}(t)\rangle = 
\langle\psi_{\bar{D}}(0)| e^{i\overline{H}^{\dagger}t} e^{-i\overline{H}t}|\psi_{\bar{D}}(0)\rangle 
\end{equation}
is not conserved in time. Since \footnote{Notice that $V(t)=e^{-i\overline{H}t}$ is a contraction operator that 
forms a strongly continuous semi-group on the projected Hilbert space $\bar{\pi}\mathcal{H}$ and satisfies 
\cite{nik_tajron2} $$\frac{d V(t)}{dt}=-i\overline{H}V(t).$$}
\begin{eqnarray}
 \partial_t \left( e^{i\overline{H}^{\dagger}t} e^{-i\overline{H}t} \right) 
&=& \left( \partial_t e^{i\overline{H}^{\dagger}t} \right) e^{-i\overline{H}t}
+ e^{i\overline{H}^{\dagger}t} \left( \partial_t e^{-i\overline{H}t} \right) 
\nonumber \\
&=& e^{i\overline{H}^{\dagger}t} i\overline{H}^{\dagger} e^{-i\overline{H}t}  
- e^{i\overline{H}^{\dagger}t} i\overline{H} e^{-i\overline{H}t} 
\nonumber \\
&=& e^{i\overline{H}^{\dagger}t} i\left( \overline{H}^{\dagger} - \overline{H} \right) e^{-i\overline{H}t}
\nonumber \\
&=& e^{i\overline{H}^{\dagger}t} \left( -\int_{\partial\bar D} d{\bf S}\cdot {\bf J} \right) e^{-i\overline{H}t} ,
\end{eqnarray}
where in the last equality we used (\ref{H-H}), we see that (\ref{norm}) implies
\begin{eqnarray}\label{norm_dot_D}
\frac{d}{dt}\langle\psi_{\bar{D}}(t)|\psi_{\bar{D}}(t)\rangle 
&=& -\langle\psi_{\bar{D}}(t)|  \int_{\partial\bar D} d{\bf S}\cdot {\bf J}({\bf x}) |\psi_{\bar{D}}(t)\rangle
\nonumber \\
&=& -\int_{\partial\bar D} d{\bf S}\cdot \langle\psi_{\bar{D}}(t)|{\bf J}({\bf x}) |\psi_{\bar{D}}(t)\rangle
\nonumber \\
&=& - \int_{\partial\bar D} d{\bf S}\cdot {\bf j}_{\bar{D}}({\bf x},t) .
\end{eqnarray} 

Note that the wave functions in (\ref{e5}) are full wave functions, not restricted wave functions. 
The full wave functions are assumed to be twice differentiable at the boundary of ${\bar{D}}$, because only in this case 
the volume integral can be turned into the surface integral via the Gauss theorem. On the other hand, 
the current ${\bf j}_{\bar{D}}$ in (\ref{norm_dot_D}) is expressed in terms of the restricted wave function $\psi_{\bar{D}}$,
which has a discontinuity at ${\bar{D}}$, implying that it is not differentiable. 
To avoid this apparent inconsistency, we must be more careful 
in specifying what we mean by integral over the ``boundary''. This really means that the surface of integration 
${\partial\bar D}$ is put {\em infinitesimally away} from the boundary towards the interior of ${\bar D}$,
where $\psi_{\bar{D}}$ coincides with $\psi$. The consequence is that ${\bf j}_{\bar{D}}$ in (\ref{norm_dot_D}) coincides 
with ${\bf j}$ defined by (\ref{curr}), implying that (\ref{norm_dot_D}) can finally be written as  
\begin{eqnarray}\label{norm_dot}
 \frac{d}{dt}\langle\psi_{\bar{D}}(t)|\psi_{\bar{D}}(t)\rangle = 
- \int_{\partial\bar D} d{\bf S}\cdot {\bf j}({\bf x},t) .
\end{eqnarray}
This shows that the rate of change of norm of the state restricted to the region ${\bar D}$
is given by the flux of the probability current through the boundary of ${\bar D}$.

%\textcolor[rgb]{0,0,1}{
Physically, the most important consequence of evolution governed by the restricted Hamiltonian (\ref{Hbar})
with an anti-hermitian boundary term 
is the change of norm of the restricted wave function, as described by (\ref{norm_dot}).
The result (\ref{norm_dot}) is rather intuitive, it can be visualized as a leak of wave function from the region 
$\bar{D}$, where the flux of the probability current quantifies how much of the wave function leaks through
the boundary of $\bar{D}$. 
%}

\section{Arrival time distribution}
\label{SECarrival}

Suppose that at the initial time $t=0$, the particle is out of the detector region $D$. This means that
\begin{equation}
 |\psi(0)\rangle =|\psi_{\bar{D}}(0)\rangle ,
\end{equation}
i.e. the initial full state is equal to the initial state restricted to ${\bar D}$. Then (\ref{norm})
is the probability $\bar{P}(t)$ that, at time $t$, the particle is in ${\bar D}$
\begin{equation}\label{a1}
 \bar{P}(t)=\langle\psi_{\bar{D}}(t)|\psi_{\bar{D}}(t)\rangle .
\end{equation}
Hence the probability that the particle is in the detector region $D$ is
\begin{equation}\label{a2}
 P(t)=1-\bar{P}(t) .
\end{equation}
Now suppose that, during a time interval $[0,T]$, the probability $P(t)$ increases with time. Then there is a positive 
function ${\cal P}(t)$ such that
\begin{equation}\label{a3}
 P(t)=\int_0^t dt'\,{\cal P}(t') ,
\end{equation} 
for any $t\in[0,T]$. This, together with (\ref{a2}), implies
\begin{equation}\label{a4}
 {\cal P}(t)=\frac{dP(t)}{dt}=-\frac{d\bar{P}(t)}{dt} .
\end{equation}
Using (\ref{a1}) and (\ref{norm_dot}), this finally gives
\begin{equation}\label{a5}
 {\cal P}(t)=\int_{\partial\bar D} d{\bf S}\cdot {\bf j}({\bf x},t) .
\end{equation}

%\textcolor[rgb]{0,0,1}{
Mathematically, the final formula (\ref{a5}) is rather compact and general. The same formula 
has also been obtained 
in \cite{nik_tajron2} by different methods, while here we derived it through the use of the 
restricted Hamiltonian (\ref{Hbar}) with the anti-hermitian boundary term.  
%}

Since $P(t)$ is a probability, it follows that ${\cal P}(t)$ in (\ref{a3}) is a probability {\em density}.
In other words, ${\cal P}(t)$ is a probability distribution. But a probability distribution of {\em what}?
We shall present two independent arguments, one heuristic and the other more rigorous, that ${\cal P}(t)$
is the probability distribution of arrival times to the detector. 

For a heuristic argument, consider first an analogous quantum equation for spatial distributions.
In a formula of the form $P=\int d^3x\, |\psi({\bf x})|^2$, the quantity $|\psi({\bf x})|^2$
is the probability density that the particle will {\em appear} at the position ${\bf x}$, rather than at any 
other position ${\bf x}'$. By analogy, ${\cal P}(t)$ in (\ref{a3})
is the probability density that the particle will {\em appear}
at the time $t$, rather than at any other time $t'$. More precisely, since $P(t)$ is the probability that the 
particle is in the detector region $D$, it follows that ${\cal P}(t)$ is the probability density that the particle will 
{\em appear} at time $t$ in the detector region $D$. The appearance of a particle in the detector at time $t$ 
means that the particle was not there immediately before $t$, so we can say that the particle {\em arrives} to detector
at time $t$. Hence we conclude that (\ref{a5}) is the {\em arrival time distribution}.

We repeat that this interpretation is only valid when $P(t)$ increases with time, i.e. when ${\cal P}(t)$ is positive.
The formula (\ref{a5}) then says that the arrival time distribution is given by the flux of the probability current
through the boundary of the detector, when the flux is positive. But what if the flux is negative? In that case
$P(t)$ {\em decreases} with time, rather than increases, so the particle departs from the detector, rather than arrives to it.
Hence, for a negative flux, the arrival probability density is zero. In this case the
$-\int_{\partial\bar D} d{\bf S}\cdot {\bf j}({\bf x},t)$ is positive and naturally interpreted as 
departure probability density \cite{{nik_tajron2}}.

Now let us confirm the conclusion above, that ${\cal P}(t)$ is arrival probability density, by a more rigorous analysis.
We first split the time interval $[0,t]$ into $k$ intervals, each of the small size $\delta t=t/k$,
and imagine that particle can only arrive at one of the times from the discrete set
$t_1=\delta t,\;t_2=2\delta t, \ldots, t_k=k\delta t=t$. At the end we shall let  $\delta t\to 0$.
Let $w(t_j)$ be the {\em conditional} probability density that the particle is in the detector at time $t_j$, 
{\em given} that it was not in the detector immediately before, at time $t_{j-1}$. 
Then the probability that it will arrive at time $t=t_k$ is
\begin{equation}\label{a29}
 {\cal P}_{\rm arr}(t)\delta t = w(t) \delta t\, \bar{P}(t-\delta t) , 
\end{equation}
where $\bar{P}(t-\delta t)$ is the probability that, at time $t-\delta t$, the particle was not in the detector region $D$
(see (\ref{a2})). But the probability $\bar{P}(t-\delta t)$ is itself a joint probability that the particle was not in $D$
at time $t-\delta t$ {\em given} that it was not there at $t-2\delta t$, 
{\em and} that it was not there at $t-2\delta t$ {\em given} that it was not there at $t-3\delta t$, etc.
Hence 
\begin{equation}\label{a30}
 \bar{P}(t-\delta t)=\prod_{j=1}^{k-1}[1-w(t_j)\delta t] ,
\end{equation}
where $1-w(t_j)\delta t$ is the conditional probability that the particle is not in $D$ 
at time $t_j$, given that it was not there at $t_{j-1}$.
Since we are interested in the limit $\delta t\to 0$, we can first write (\ref{a30}) as 
\begin{eqnarray}
 \bar{P}(t-\delta t) &=& \prod_{j=1}^{k-1} \exp \left(-w(t_j)\delta t +{\cal O}(\delta t^2) \right)
\nonumber \\ 
&=& \exp \left(-\displaystyle\sum_{j=1}^{k-1} [w(t_j)\delta t +{\cal O}(\delta t^2)] \right),
\end{eqnarray}
and then take the limit $\delta t\to 0$, which gives
\begin{equation}
  \bar{P}(t-\delta t) = \bar{P}(t) = e^{-\int_0^t dt'\,w(t')} .
\end{equation}
Thus (\ref{a29}) in the limit $\delta t\to 0$ can be written as 
\begin{eqnarray}
 {\cal P}_{\rm arr}(t) &=& w(t)\bar{P}(t)=w(t)e^{-\int_0^t dt'\,w(t')}
\nonumber \\
&=& -\frac{d\bar{P}(t)}{dt}= {\cal P}(t) ,
\end{eqnarray}
where in the last equality we used (\ref{a4}). This shows that (\ref{a5}) is indeed the arrival probability density,
provided that it is positive.

%\textcolor[rgb]{0,0,1}{
The measurable predictions of the arrival time distribution based on flux of the probability current can in principle 
be distinguished experimentally from predictions of the arrival time distribution based on other approaches.
It is beyond the scope of the present paper to discuss such measurable differences in detail, but they have
been studied elsewhere \cite{maccone}.  
%}

\section{Summary and conclusion}
\label{SECconclusion}

The results of this paper can be summarized as follows.
As the wave function of a particle approaches the detector, 
a part of the full wave function leaks into the detector 
region $D$, so the other part of wave function, that remains outside of $D$, diminishes with time.
Since the norm of full wave function $\psi({\bf x},t)$ does not depend on time, 
the norm of its restriction $\psi_{\bar{D}}({\bf x},t)$
to the region $\bar{D}$ outside of the detector
depends on time. 
Therefore the ``Hamiltonian" $\overline{H}$ governing the time evolution of the restricted 
wave function $\psi_{\bar{D}}({\bf x},t)$ must be a non-hermitian operator.
In this paper we have found an explicit representation of $\overline{H}$ and found that its
non-hermitian part can be written as a boundary term, proportional to the flux 
of the probability current operator through the boundary $\partial\bar{D}$ of $\bar{D}$. 
The explicit representation of $\overline{H}$ is given by Eq.~(\ref{Hbar}).
From the time-dependent norm of $\psi_{\bar{D}}({\bf x},t)$ we have computed the 
arrival time probability density, namely, the probability that the particle will be 
detected to arrive to the detector between the times $t$ and $t+dt$, and found 
that this arrival probability density is equal to the flux of the probability current 
through the boundary $\partial\bar{D}$. 

Our final result, that the arrival probability density 
is equal to the flux of the probability current, has also been
obtained by other approaches, based on standard QM 
\cite{kijowski,delgado-muga,tumulka1,nik_tajron2}, 
as well as on Bohmian particle trajectories \cite{leavens,leavens_book,durrtime1,durrtime2}.
The approach of the present paper, also based on standard QM, is complementary to the existing approaches,
because we arrived to the same conclusion by using different methods.
However, we stress that it is not generally 
accepted in the literature that the arrival probability density should be equal
to the flux of the probability current, see \cite{muga-physrep,V1.10,V2.4,maccone} for reviews 
of other proposals, so we believe that the result of this paper
is a valuable contribution towards a resolution of an important problem in physics.

%\textcolor[rgb]{1,0,0}{In this paper we have presented a new method of finding quantum probability density of arrival 
%at the detector. The main result is that the evolution of the restricted wave function is governed 
%by the non-hermitian operator $\overline{H}$. The non-hermiticity of $\overline{H}$ is due to its non-triviality 
%at the boundary of $\bar{D}$. This results in a existence of a boundary term that is proportional to the flux of the 
%probability current operator $\bf{J}$, and further leads to a non-conservation of the norm of the restricted wave function 
%$\psi_{\bar{D}}$. }\\

%\textcolor[rgb]{1,0,0}{Komentirati veze i razlike s našim/tvojim prijasnjim radovima, 
%ali i usporedbu/razlike s drugim autorima (Tumulka, etc?)}

\section*{Acknowledgments}

%This work was supported  
%by the Ministry of Science of the Republic of Croatia.
The work of T.J. is supported by Croatian Science
Foundation Project No. IP-2020-02-9614.

% \section*{Statements and Declarations}
% % OBAVEZNO ZA EPJ+
% 
% Data sharing not applicable to this article as no datasets were generated or analyzed during the current study 
% and article describes entirely theoretical research.

\appendix

\section{The adjoint $\overline{H}^\dagger$ and non-hermiticity of $\overline{H}$}

To extract the non-hermitian part of $\overline{H}$ it is sufficient to find its adjoint $\overline{H}^\dagger$. 
Notice that while obtaining \eqref{Hbar1_dagger} we used the na\"{i}ve rule for calculating the adjoint of product of operators, 
$(AB)^\dagger=B^\dagger A^\dagger$, which is valid only in finite dimensional spaces or for some particular set of wave functions. 
Here we will show that the non-hermitian part of $\overline{H}$ can be extracted directly from its adjoint in a straightforward 
and rigorous way once the proper definition of the adjoint is used. 

We start with the definition of adjoint \cite{reed, moretti}:

\textit{The adjoint $A^\dagger : {\cal D}(A^\dagger)\longrightarrow \mathcal{H}$ 
(with ${\cal D}$ denoting the domain of the operator)
of a densely defined linear operator 
$A:{\cal D}(A)\longrightarrow\mathcal{H}$ 
is defined by
\begin{itemize}
\item ${\cal D}(A^\dagger):=$\\
$\left\{\psi\in\mathcal{H}\  | \ 
\exists\eta\in\mathcal{H}:\forall\alpha\in {\cal D}(A):\langle\psi|A\alpha\rangle=\langle\eta|\alpha\rangle\right\}$
\item  $A^\dagger \psi=\eta$.
\end{itemize}}

Then we analyze the expression in \eqref{e3} and notice the following chain of equalities
\begin{equation}\begin{split}\label{aa1}
\langle\psi|\overline{H}|\varphi\rangle &\equiv \int_{\bar D} d^3x \, \psi^*({\bf x}) 
\left[  -\frac{\mbox{\boldmath $\nabla$}^2}{2m}+V({\bf x}) \right] \varphi({\bf x})\\
 &=\int_{\mathbb{R}^3} d^3x \, \psi^*({\bf x}) 
\chi_{\bar D}({\bf x}) \hat{H}\varphi({\bf x})\\
&=\langle\psi|\bar{\pi}H|\varphi\rangle,
\end{split}
\end{equation}
where $\chi_{\bar D}({\bf x})$ is the characteristic function of the region ${\bar D}$
\begin{equation}
 \chi_{\bar D}({\bf x}) =\left\{ 
\begin{array}{cl}
 \displaystyle 1 & \;\; {\rm for} \;\; {\bf x}\in \bar{D}  \\
 0 & \;\; {\rm for} \;\; {\bf x}\notin \bar{D} ,
\end{array}
\right.
\end{equation}
and $\bar{\pi}$ is defined in (\ref{barpi}).
This means that
\begin{equation}
 \langle{\bf x}|\overline{H}|\varphi\rangle =\langle{\bf x}|\bar{\pi}H|\varphi\rangle= \chi_{\bar{D}}({\bf x})\hat{H} 
\varphi({\bf x}),
\end{equation} 
where $\varphi\in {\cal D}(H)$ and $\psi\in\mathcal{H}$. Now, in order to obtain the adjoint $\overline{H}^\dagger$ we need 
to derive $\eta$ from the general definition for the adjoint. This will enable us to find both the domain and the 
``rule of action'' of the operator $\overline{H}^\dagger$.   For that matter we will use the simple identity 
\begin{equation}\label{ident}
\psi^*\mbox{\boldmath $\nabla$}^2\varphi=\mbox{\boldmath $\nabla$}
\left[\psi^*\mbox{\boldmath $\nabla$}\varphi-(\mbox{\boldmath $\nabla$}\psi^*)\varphi\right]
+(\mbox{\boldmath $\nabla$}^2\psi^*)\varphi
\end{equation}
in order to rewrite \eqref{aa1}  as
\begin{equation}\begin{split}\label{aa3}
\langle\psi|\overline{H}|\varphi\rangle&=\int_{\bar D} d^3x \, 
\left(  -\frac{\mbox{\boldmath $\nabla$}^2\psi^*({\bf x})}{2m}+V({\bf x})\psi^*({\bf x}) \right)\varphi({\bf x})\\
&-\frac{1}{2m}\int_{\bar D} d^3x \, \mbox{\boldmath $\nabla$}
\left[\psi^*\mbox{\boldmath $\nabla$}\varphi-(\mbox{\boldmath $\nabla$}\psi^*)\varphi\right].
\end{split}\end{equation}
The first term is equal to
\begin{equation}\begin{split}\label{prvi}
&\int_{\bar D} d^3x \, \left(  -\frac{\mbox{\boldmath $\nabla$}^2\psi^*({\bf x})}{2m}+V({\bf x})\psi^*({\bf x}) 
\right)\varphi({\bf x})   \\
&=\int_{\mathbb{R}^3} d^3x \, \left( \chi_{\bar{D}}\hat{H}\psi \right)^* \varphi\\
&=\langle \overline{H}\psi|\varphi\rangle ,
\end{split}
\end{equation}
while for the second term we have
\begin{equation}\begin{split}\label{drugi}
&-\frac{1}{2m}\int_{\bar D} d^3x \, \mbox{\boldmath $\nabla$}\left[\psi^*\mbox{\boldmath $\nabla$}\varphi
-(\mbox{\boldmath $\nabla$}\psi^*)\varphi\right]    \\
&=-\frac{1}{2m}\int_{\partial\bar D} d{\bf S}\cdot \left[\psi^*\mbox{\boldmath $\nabla$}\varphi-(\mbox{\boldmath $\nabla$}\psi^*)
\varphi\right]\\
&=\frac{i}{2m}\int_{\partial\bar D} d{\bf S}\cdot\left(\langle\psi|{\bf x}\rangle\langle {\bf x}|{\bf p}\varphi\rangle 
+ \langle\psi|{\bf p}|{\bf x}\rangle\langle {\bf x}|\varphi\rangle\right)\\
&=\langle\psi| \frac{i}{2m}\int_{\partial\bar D} d{\bf S}\cdot\left\{|{\bf x}\rangle\langle {\bf x}|, {\bf p} \right\}  
|\varphi\rangle \\
&=\langle i\int_{\partial\bar D} d{\bf S}\cdot \bf{J} \psi |\varphi\rangle, 
\end{split}
\end{equation}
where we have used the Gauss theorem and the definitions \eqref{deff1} and \eqref{deff2}, 
the $\partial\bar D$ is the boundary of $\bar D$, and $d{\bf S}$ 
is the area element directed outwards from $\bar D$.
Here we have to notice that the identity \eqref{ident} is only valid if $\psi$ is at least a function of class 
$\mathcal{C}^2(\mathbb{R}^3)$ and therefore the equation \eqref{aa3} is well defined only if $\psi\in\mathcal{H}$ is such that 
$\mbox{\boldmath $\nabla$}^2\psi\in\mathcal{H}$, which will define the domain of the adjoint ${\cal D}(\overline{H}^\dagger)$. 
Equations \eqref{aa3}, \eqref{prvi} and \eqref{drugi} together lead to
\begin{equation}\begin{split}
\langle\psi|\overline{H}\varphi\rangle&=\langle\left(\overline{H}
+i\int_{\partial\bar D} d{\bf S}\cdot \bf{J} \right)\psi|\varphi\rangle\\
&=\langle\overline{H}^\dagger\psi|\varphi\rangle
\end{split}
\end{equation}
and we can explicitly read out the ``rule of acting'' for the adjoint operator $\overline{H}^\dagger$
\begin{equation}\label{aafin}
\overline{H}^\dagger=\overline{H}+i\int_{\partial\bar D} d{\bf S}\cdot \bf{J} ,
\end{equation}
which, together with the domain ${\cal D}(\overline{H}^\dagger)$, fully defines the operator $\overline{H}^\dagger$. 
Equation \eqref{aafin} explicitly shows the non-hermiticity of $\overline{H}$ and is in complete agreement with Eq.~\eqref{H-H}.

The non-hermiticity of $\overline{H}$ can also be described by the operator
\begin{equation}
 N \equiv i(\overline{H}-\overline{H}^\dagger)=\int_{\partial\bar D} d{\bf S}\cdot \bf{J} .
\end{equation}
This operator is related to the arrival time distribution \eqref{a5} via its expectation value
\begin{equation}
 \mathcal{P}(t)=\langle\psi(t)|N|\psi(t)\rangle=\int_{\partial\bar D} d{\bf S}\cdot {\bf j}({\bf x},t) ,
\end{equation}
which can also be written as a volume integral 
\begin{equation}
\begin{split}
\mathcal{P}(t)&=\int_{\bar D} d^3x\,  \mbox{\boldmath $\nabla$} \cdot {\bf j}({\bf x},t) \\
&=\int_{\mathbb{R}^3} d^3x\,\chi_{\bar D}({\bf x})  \mbox{\boldmath $\nabla$} \cdot {\bf j}({\bf x},t) \\
&=-\int_{\mathbb{R}^3} d^3x   \left( \mbox{\boldmath $\nabla$}\chi_{\bar D}({\bf x}) \right) \cdot {\bf j}({\bf x},t) .
\end{split}
\end{equation}

\mbox{}

\end{document}